\documentclass[12pt]{article}
\begin{document}
\begin{center}

{\huge \bf A Version of Exclusively Perturbative Quantum Field Theory}
 \\[10mm]
 S.A. Larin \\ [3mm]
 Institute for Nuclear Research of the
 Russian Academy of Sciences,   \\
 60-th October Anniversary Prospect 7a,
 Moscow 117312, Russia
\end{center}

\vspace{30mm}
\begin{abstract}
We suggest a version of renormalizable Quantum Field Theory which does not contain
non-perturbative effects.
This is otained by the proper use of the boundary conditions in the
functional integral of the generating functional of Green functions.
It is well known which
boundary conditions are applied to the fields
of the functional integral to get correct perturbation theory.
We propose  that these  conditions should be used for all
fields integrated in the generating functional integral.
It is shown that in this case non-perturbative effects are absent.
That is we assume that perturbation theory defines the complete
generating functional integral. It allows, in particular,
to formulate the generating functional integral in a unique way
as an exact compact mathematical formula.

\end{abstract}

\newpage
\section{Introduction}
Most of precision tests of Quantum Field Theory are performed within
perturbative approach. The famous example of this type is the anomalous
magnetic moment of the electron. The agreement between the  Quantum Electrodynamics
prediction and the experiment in this case 
within at least ten decimal points \cite{pdg}
convinces us in the validity of renormalizable Quantum Field Theory.

The known example of applcations of non-perturbative effects is the QCD
sume rule method \cite{svz}, where hadronic properties are calculated
by means of the so called non-perturbative quark and gluon condensates.
As it will be discussed below these condensates can be, in principle,
generated within perturbation theory.

Quite recently a solution to the strong CP-problem based
on the proper definition of the integration region in the functional integral
for the generating functional of Green functions of
Quantum Chromodynamics was suggested in the work \cite{l}. 
This solution defines that
the integration measure includes only fields decreasing at infinity, excluding 
all other fields like instantons \cite{inst}.
There is the statement \cite{h},\cite{h1} that instantons solve
the known  $U(1)$ problem. Thus
 it seems that they should not be discarded
from the theory. But  there is also 
the known solution \cite{ks}
to the  $U(1)$ problem applying the axial anomaly
which was suggested before the discovery of instantons.

In the present paper we suggest a version of Quantum Field Theory
without non-perturbative contributions
and, in particular, demonstrate that the solution \cite{l} to the
strong CP problem implies that
the corresponding theory is free of non-perturbative effects.

\section{Main Part}
Let us consider a functional integral for the generating functional
of Green functions of some generic theory (this can be QCD or
QED or any other theory):

\begin{equation}
\label{gf}
Z(J)=\frac{1}{N}\int d\Phi~~ exp\left(i\int d^4x\left( L_0(x) 
+gL_{int}(x) +J_k(x)\cdot \Phi_k(x) 
\right)\right), 
\end{equation}
where $d\Phi$ is the integration measure of the functional integral
$Z(J)$ including all allowed types of the fields $\Phi_k$ of the theory.
$J_k$ are sources of these fields. $J$ in $Z(J)$ is
the full set of $J_k$. $L_0(x)$ is the free Lagrangian
and $gL_{int}(x)$ is the interaction Lagrangian of the considered theory.
$g$ is the coupling constant of interacrions of the theory.
$N$ is the standard normalization factor.
Generalization to the case of several coupling constants of interactions
is completely straightforward.

Of course the generating functional (\ref{gf}) is not completely defined
by the defining the Lagrangiann $L(x)= L_0(x) +gL_{int}(x)$ only.
One should also define which types of the fields $\Phi_k$ are allowed
in the integration measure. If one allows integration over all possible
configurations of the fields $\Phi_k$, then it is not possible to reproduce
the perturbative fields propagators of the  necessary forms, i.e. 
with the correct '$i\epsilon$'
prescription of the type $1/(k^2-m^2 +i\epsilon)$.

To obtain the perturbative propagators of the correct forms one should impose
on the integrated fieelds the known boundary conditions.
For example, one has the following boundary conditions
for the gluon fields in QCD:
\begin{equation}
\label{bou}
A_{\mu}^a(\vec{x},t \to  \infty) \to A_{\mu,in}^a(x),
\end{equation}
\[
A_{\mu}^a(\vec{x},t \to  \infty) \to A_{\mu}^{a,out}(x),
\]
where the incoming  gluon  fields $A_{\mu,in}^a(x)$ contain only
the positive frequency part and, in opposite, the  
outgoing asymptotic gluon  fields $A_{\mu,}^{a,out}(x)$
contain the negative frequency part:

\begin{equation}
\label{boun}
A_{\mu,in}^a(x)=\frac{1}{(2\pi)^{3/2}} \int d^3 k~ e^{i(\vec{k}\vec{x}-\omega t)}
v_{\mu}^i(k)a_i(k)/\sqrt{2w}, \\
\end{equation}
\[
A_{\mu,}^{a,out}(x)=\frac{1}{(2\pi)^{3/2}} \int d^3 k~
e^{i(\vec{k}\vec{x}+\omega t)}
v_{\mu}^i(k)a^*_i(k)/\sqrt{2w}.
\]
Here $\omega=\sqrt{\vec{k}^2}$, $v_{\mu}^i(k)$  are
the polarization vectors of the gluons. In (\ref{boun}) sums over
gluon polarizations $i=1,2$ are assumed.

These are the known Feynman boundary conditions of emission. They
ensure the correct forms of perturbative propagators
of the fields
of the type $1/(k^2+i\epsilon$) with the necessary 
plus $i\epsilon$ prescription, see \cite{sf}.

Hence the gluon fields (the quark fields also) in QCD
with emission boundary conditions
 oscillate at time infinities. Making transition to the 
Euclidean space with the help of the Wick rotation $t\rightarrow ix_4$ one gets that
 the fields decrease at  time infinities. Thus it is easy to see 
that in the Euclidean space 
total derivatives in the Lagrangian are zero within perturbation theory.
This allows, in particular, to solve \cite{l} the Strong CP problem
without involving hypothetical exotic particles like axions; 
for the review of axions
see \cite{gs}.

One can  write perturbative boundary conditions (\ref{bou}) for all fields
$\Phi_i$ of the  considered theory (\ref{gf}) symbolically as follows:
\begin{equation}
\label{bou1}
\Phi(t \to \pm\infty)\to\Phi_{in}^{out}(x).
\end{equation}
This short notation is used below to formulate the functional intergal
for the generating functional of Green functions 
as a uniquely defined compact  formula.

If one defines the generating functional with integrations
over all possible field configurations, then one should also
integrate over the non-perturbative fields like instantons \cite{inst}
in addition to the perturbative fields with emission boundary conditions.
These non-perturbative fields produce only non-perturbative contributions and do not effect
the perturbative propagators. But then one comes to uncertainties
since there can be, in principle, other non-perturbative  solutions like instantons
and it is not known in advance how many of them exist.

That is why it seems to be natural to use the  boundary
conditions of emission (\ref{bou1}) for all
fields $\Phi_i$  over which the integration in the the functional
integral (\ref{gf}) proceeds. Then, in particular, all fields decrease in the
Euclidean space at the time infinities. Hence 
total derivatives in the Lagrangian are nullified and 
this, in particular, solves \cite{l} the strong CP problem.
Besides, this definition of the boundary conditions  allows to formulate 
uniquely the complete  generating functional
integral of Green functions of the theory as an exact compact mathematical formula:
\begin{equation}
\label{gf1}
Z(J)=\frac{1}{N}\int_
{\Phi(t \to \pm\infty)\to\Phi_{in}^{out}}
 d\Phi~exp\left(i\int d^4x\left( L_0(x)+gL_{int}(x) +J_k\cdot \Phi_k
\right)\right).
\end{equation}

It turns out that this expression for the  generating functional
integral contains only perturbative contributions
and does not contain non-perturbative ones.
Let us demonstrate it.

One can straightforwardly expand the generating functional (\ref{gf1})
in the perturbative series in the coupling constant $g$ at the 
point $g=-0$ (in the same way as one expands
at the point $g=+0$). The expansion is well defined. 
This is the perturbation theory with the finite
coefficients (the counterterms are assumed to be included into the Lagrangian)
since integrations in the functional integral (\ref{gf1}) proceed only over the fields
with perturbative boundary conditions of emission. It excludes the
presence of the  non-perturbative
contributions of the type $exp(-1/g)$ since otherwise the expansion 
would produce infinities at the point $g=-0$.

One can also consider an expansion of the functional integral (\ref{gf1})
at the point $g=i0$. Again the coefficients of the expansion will be finite
on the same reasons as above. This excludes the presence of the non-perturbative
terms of the type $exp(-1/g^2)$ since infinities would be generated otherwise.

More generally, to demonstrate the absence of the non-perturbative terms
of the type $exp(-1/g^x)$, where $x$ is an arbitrary positive number, 
one can consider an expansion at the
point  $g=i^{2/x} 0$. The expansion is finite.
This demonstrates the absence of the non-perturbative contributions
in the generating functional (\ref{gf1}).

There is the question of the so called non-perturbative
vacuum quark and gluon condensates used in the method of the QCD
sum rules \cite{svz} for calculation of hadronic properties.
These condensates have the values 

$<\bar{q}q>\approx (200MeV)^3$
and $\alpha_s<G_{\mu\nu}^aG_{\mu\nu}^a>\approx (400 MeV)^4$
at the normalization point of the order of $1GeV$.
Here $q$ is the quark field, $G_{\mu\nu}^a$ is the gluon strength
tensor, $\alpha_s=g_s^2/4\pi^2$, where $g_s$ is the strong
coupling constant in the QCD Lagrangian.

But, in principle, one can generate these condensates within perturbation
theory with massive perturbative  propagators after the summations
of the complete asymptotic perturbative series for the corresponding
vacuum expectation values.

\section{Conclusions}
We suggest a version of Quantum Field Theory which does not contain
non-perturbative effects.
This is otained by the proper, in our opinion, use of the boundary conditions in the
functional integral of the generating functional of Green functions.
It is well known which
boundary conditions are applied to the fields
of the functional integral to get correct perturbation theory.
These are the boundary conditions which ensure that the fields decrease
at the time infinities nullifying all total derivatives in the Lagrangian
of the theory.
We propose  that these  conditions should be used for all
fields integrated in the generating functional integral.
It is shown that in this case the non-perturbative effects are absent.
That is we assume that perturbation theory defines the complete
generating functional integral. It allows, in particular, 
to formulate the generating functional integral in a unique way
as an exact compact mathematical formula.

\section{Acknowledgments}
The author is grateful
to the members of the Theory Division of INR
for helpful discussions.

\end{document}